\documentclass[12pt]{amsart}

\usepackage{amsfonts,amssymb,amsmath,amscd}
\usepackage{bbm}
\usepackage[ps,dvips,matrix,arrow,frame,import,curve,color]{xy}
\newcommand{\ra}{\rightarrow}

\newcommand{\Tr}{{\rm Tr}}

\newcommand{\ZZ}{{\mathbb Z}}
\newcommand{\RR}{{\mathbb R}}
\newcommand{\CC}{{\mathbb C}}

\newcommand{\tpsi}{{\tilde\psi}}
\newcommand{\teta}{{\tilde\eta}}

\newcommand{\cO}{{\mathcal O}}
\newcommand{\bpartial}{{\bar\partial}}
\newcommand{\eps}{\epsilon}

\newcommand{\ad}{{\rm ad}}

\renewcommand{\Re}{{\rm Re}}
\renewcommand{\Im}{{\rm Im}}
\newcommand{\bz}{{\bar z}}

\newcommand{\cA}{{\mathcal A}}

\newcommand{\Hom}{{\rm Hom}}

\newcommand{\tq}{{\tilde q}}

\newcommand{\cM}{{\mathcal M}}

\newcommand{\End}{{\rm End}}

\newcommand{\tp}{{\tilde p}}

\newcommand{\LG}{{{}^LG}}
\newcommand{\Ltau}{{{}^L\tau}}

\newcommand{\bsigma}{{\bar\sigma}}
\newcommand{\btau}{{\bar\tau}}
\newcommand{\MH}{{{\mathcal M}_H}}
\newcommand{\frt}{{\mathfrak t}}

\newcommand{\Bun}{{\rm Bun}}

\newcommand{\frg}{{\mathfrak g}}
\newcommand{\cL}{{\mathcal L}}
\newcommand{\Lt}{{{}^L t}}

\newcommand{\tU}{{\tilde U}}
\newcommand{\bbA}{{\mathsf A}}
\newcommand{\qq}{{\mathsf q}}
\newcommand{\Lqq}{{{}^L{\mathsf q}}}
\newcommand{\dch}{{h^\vee}}
\newcommand{\bbT}{{\mathbb T}}
\newcommand{\Lalpha}{{{}^L\alpha}}
\newcommand{\Leta}{{{}^L\eta}}

\title[Quantum Geometric Langlands]{A Note on Quantum Geometric Langlands Duality, Gauge Theory, and
Quantization of the Moduli Space of Flat Connections}

\author{Anton Kapustin}
\email{kapustin@theory.caltech.edu}
\address{California Institute of Technology}

\begin{document}

\begin{titlepage}

\maketitle

\begin{abstract}
Montonen-Olive duality implies that the categories of A-branes on
the moduli spaces of Higgs bundles on a Riemann surface $C$ for
groups $G$ and $\LG$ are equivalent. We reformulate this as a
statement about categories of B-branes on the quantized moduli
spaces of flat connections for groups $G_\CC$ and $\LG_\CC$. We show
that it implies the statement of the Quantum Geometric Langlands
duality with a purely imaginary ``quantum parameter'' which is
proportional to the inverse of the Planck constant of the gauge
theory. The ramified version of the story is also considered.

\end{abstract}

%\vspace{-7in}
%\parbox{\linewidth}
%{\small\hfill \shortstack{CALT-xx-xxxx}} \vspace{6in}

\end{titlepage}

\section{Introduction}

Langlands duality has many different manifestations in mathematics
and physics. The most famous examples in mathematics are the
Langlands program for number fields and function fields and its
geometric counterpart. It has been proposed a long time ago by
M.~Atiyah that Langlands Program is related to Montonen-Olive
duality in gauge theory. Montonen-Olive duality \cite{MO}
(formulated more precisely by Osborn \cite{Os}) is a conjecture that
four-dimensional $N=4$ supersymmetric gauge theories with gauge
groups $G$ and $\LG$ are isomorphic if their coupling constants are
inversely related. Here $G$ and $\LG$ are compact simple Lie groups
such that the character lattice of $G$ is isomorphic to the
cocharacter lattice of $\LG$ and vice versa. In \cite{KW} it was
shown how the main statements of the Geometric Langlands Program in
the unramified case can be deduced by considering a topologically
twisted version of the Montonen-Olive duality. This was later
extended to the ramified case
\cite{GW:ramified,Witten:wild,GW:rigid}.

Geometric Langlands duality has a ``baby'' version which says that
the derived categories of coherent sheaves of the moduli spaces of
Higgs bundles over a Riemann surface $C$ for a group $G$ and its
Langlands-dual $\LG$ are equivalent. This statement can be deduced
by applying the Montonen-Olive duality to a holomorphic-topological
twist of the $N=4$ supersymmetric gauge theory \cite{Kap:HT}.

Yet another mathematical manifestation of Langlands duality is the
Quantum Geometric Langlands (QGL) duality which relates the
categories of twisted D-modules over moduli spaces (actually,
stacks) of principal $G_\CC$ and $\LG_\CC$ bundles over $C$. We
denote these moduli spaces $\Bun_G(C)$ and $\Bun_\LG(C)$. QGL is a
generalization of the usual Geometric Langlands duality, in the
sense that in a certain ``classical'' limit the former reduces to
the latter.\footnote{Confusingly, the ``baby'' version of the
Geometric Langlands duality mentioned in the previous paragraph is
sometimes referred to as the classical limit of the Geometric
Langlands duality. We will not use this terminology in this paper.}
For a review of QGL and its relation to Conformal Field Theory see
\cite{Frenkel}; for some physical manifestations of QGL see
\cite{RT,three}.

A twisted D-module over a complex manifold $M$ is a module over the
sheaf of (holomorphic) differential operators on a complex power of
a holomorphic line bundle $\cL$ over $M$. In the QGL context $\cL$
is a certain line bundle $\cL_0$ over $\Bun_G(C)$,\footnote{For
$G=SU(N)$ $\cL_0$ is the determinant line bundle; in general it is
determined by the property that its first Chern class generates the
second cohomology of $\Bun_G(C)$.} so twisted D-modules are modules
over the algebra of differential operators on $\cL_0^{\qq-\dch}$,
$\qq\in\CC$. Here $\dch$ is the dual Coxeter number of $G$. Note
that $\qq=0$ corresponds not to untwisted D-modules, but to modules
over holomorphic differential operators on the line bundle
$\cL_0^{-\dch}$; this shift in parametrization is introduced for
future convenience. We will refer to the twist by $\cL_0^{-\dch}$ as
the critical twist.

According to QGL, the derived categories of twisted D-modules on
$\Bun_G(C)$ and $\Bun_\LG(C)$ are equivalent if the twist parameters
$\qq$ and $\Lqq$ are related by
\begin{equation}\label{Psireal}
\Lqq=-\frac{1}{n_\frg\qq},
\end{equation}
where $n_\frg =1,2,$ or $3$ depending on the maximal multiplicity of
edges in the Dynkin diagram of $G$. If we interpret $\qq$ as the
inverse of a ``Planck constant'', then the ``classical'' limit
$\qq\ra \infty$ on one side is dual to the ``ultra-quantum'' limit
$\Lqq\ra 0$ on the other side. (We will see shortly why it is
natural to identify $\qq$ as the inverse of the Planck constant.) In
this limit QGL duality reduces to classical geometric Langlands
duality. Namely, one gets the statement that the derived category of
coherent sheaves on the moduli space of flat $G_\CC$ connections
(which is regarded as the $\qq\ra\infty$ limit of the derived
category of twisted D-modules on $\Bun_G(C)$) is equivalent to the
derived category of critically twisted D-modules on $\Bun_\LG(C)$.

In \cite{KW} QGL duality was interpreted in terms of equivalence of
categories of A-branes on the Hitchin moduli spaces $\MH(G,C)$ and
$\MH(\LG,C)$ with respect to a certain complex structure $K$ and
with a nonzero B-field. The B-field is determined by a gauge theory
parameter $\theta$ which takes values in $\RR$. It was argued in
\cite{KW} that the category of A-branes on $\MH(G,C)_K$ is
equivalent to the category of twisted D-modules on $\Bun_G(C)$ with
the twist parameter
$$
\qq=\frac{\theta}{2\pi}.
$$
Montonen-Olive duality implies that the categories of A-branes for
$G$ and $\LG$ are equivalent if $\qq$ and $\Lqq$ are related as in
(\ref{Psireal}); this implies the statement of the QGL duality for
real $\qq$.

One drawback of this physical derivation is that $\qq$ is naturally
real, while in QGL it is complex. Another somewhat unsatisfactory
feature is that from the physical viewpoint $\theta$ is unrelated to
the Planck constant of the underlying gauge theory; the latter does
not appear because it affects only the BRST exact terms in the gauge
theory action.

In this note we describe a variation of the argument of \cite{KW}
which provides an alternative derivation of the QGL duality from
gauge theory. Our starting point will again be the dual pair of
$N=4$ supersymmetric gauge theories with gauge groups $G$ and $\LG$.
We will twist them into a dual pair of topological field theories as
in \cite{KW}. Recall that each of these topological field theories
has a 1-parameter family of BRST operators labeled by
$t\in\CC\bigcup \{\infty\}$. The Montonen-Olive duality acts on this
family by
$$
t\ra \Lt= t \frac{|\tau|}{\tau},
$$
where $\tau$ is the complexified gauge coupling of the $N=4$ gauge
theory:
$$
\tau=\frac{\theta}{2\pi}+\frac{4\pi i}{e^2}.
$$
Classical Geometric Langlands duality is related to the special case
$\theta=0$, $t=i,\Lt=1$. In this note we look at the case $t=\Lt=0$,
with arbitrary $\theta$ and $e^2$. It will be shown below that for
$\theta=0$ the relevant category is the derived category of twisted
D-modules on $\Bun_G(C)$ with $\qq=\btau$. Thus the parameter $\qq$
is now purely imaginary, and its imaginary part is precisely the
inverse of the gauge-theory Planck constant. The limit $\qq\ra
\infty$ is the classical limit in the gauge theory, so classical
Geometric Langlands duality emerges as one of the two dual theories
becomes classical.

The interpretation in terms of branes also changes: instead of
A-branes in complex structure $K$, we will obtain A-branes in
complex structure $I$, with respect to which $\MH(G,C)$ can be
identified with the moduli space of Higgs bundles on $C$. This
interpretation makes sense for arbitrary $\theta$. The connection
with twisted D-modules arises through an analogue of the canonical
coisotropic brane of \cite{KW}; unlike in \cite{KW}, this analogue
exists only for $\theta=0$. As explained below, yet another
reformulation of the QGL is as a statement of
derived-Morita-equivalence between the quantized moduli spaces of
flat connections for $G_\CC$ and $\LG_\CC$.

It is interesting to note that Goncharov and Fock \cite{GF}
considered the quantization of certain moduli spaces related to the
moduli space of flat connections and found signs of Langlands
duality. However, the details of their set-up are rather different
from ours and we do not understand the precise relationship.

An important role in the Geometric Langlands duality is played by
Wilson and 't Hooft line operators. For $t=0$ and $t=\infty$ we
argue that there are no such line operators in the topologically
twisted gauge theory. On the other hand, there are still interesting
surface operators defined in \cite{GW:ramified}, as well as line
operators bound to surface operators. We discuss how the inclusion
of surface operators leads to a ramified version of Quantum
Geometric Langlands duality.

The author is grateful to Alexander Goncharov and David Ben-Zvi for
discussions. This work was conceived while attending a conference
"Gauge Theory and Representation Theory" at the Institute for
Advanced Study, Princeton, and completed at the Kavli Institute for
Theoretical Physics, Santa Barbara. The authors would like to thank
these institutions for hospitality. This work was supported in part
by the DOE grant DE-FG03-92-ER40701 and DARPA and AFOSR through the
grant FA9550-07-1-0543.

\section{GL-twisted theory at $t=0$}

We use the notation of \cite{KW} and \cite{GW:ramified} throughout.
The fields of the GL-twisted theory are a connection $A$ on a
principal $G$-bundle $E$, a 1-form $\phi$ with values in $\ad(E)$, a
section $\sigma$ of $\ad(E)_\CC$, a pair of 1-forms $\psi,\tpsi$
with values in $\ad(E)_\CC$, a pair of sections $\eta,\teta$ of
$\ad(E)_\CC$, and a 2-form $\chi$ with values in $\ad(E)_\CC$. The
fields $A,\phi,\sigma$ are bosonic, while
$\psi,\tpsi,\eta,\teta,\chi$ are fermionic. As in \cite{KW}, we use
the convention that the covariant derivative is $d+A$; thus in a
unitary trivialization $A$ and $\phi$ are anti-Hermitian.

The GL-twisted theory has two supercommuting BRST operators $Q_\ell$
and $Q_r$ originating from the left-handed and right-handed
supersymmetries of the $N=4$ gauge theory. The most general BRST
operator is therefore
$$
Q_t=Q_\ell+t Q_r, \quad t\in\CC\cup \{\infty\}.
$$
In this paper we will study the topological theory at $t=0$. This is
the case when the BRST operator is ``chiral'', i.e. is purely
left-handed. The BRST transformations take the form
\begin{align*}
\delta A &=\psi, & \delta\phi &=-i\tpsi,\\
\delta \sigma &=0, & \delta\bsigma &=i\eta,\\
\delta\eta &=[\bsigma,\sigma], & \delta\teta &= - D*\phi,\\
\delta\psi &= D\sigma, & \delta\tpsi &= -[\phi,\sigma], \\
\delta\chi
&=\left(F-\phi\wedge\phi\right)^+ - \left(D\phi\right)^-. & &
\end{align*}
The theory has a $U(1)$ symmetry (ghost-number symmetry) with
respect to which $\sigma$ has charge $2$, $\psi,\tpsi$ have charge
$1$, $\eta,\teta,\chi$ have charge $-1$, and $A,\phi$ have charge
$0$.

The only local BRST-invariant and gauge-invariant observables in the
theory are gauge-invariant polynomials built out of $\sigma$. Since
they have positive ghost-number charge, all correlators of these
operators vanish. We will see below the theory has no interesting
line operators, but it has surface operators as well as a variety of
boundary conditions.

On a compact manifold without boundary and without insertion of
nonlocal operators, one may consider computing the partition
function. The action can be written in the form
$$
S=\{Q,V\}+\frac{i\btau}{4\pi} \int_M \Tr\, F\wedge F,
$$
where $\btau$ is the complex-conjugate of
$$
\tau=\frac{\theta}{2\pi}+\frac{4\pi i}{e^2}.
$$
Thus the partition function is an anti-holomorphic function of
$\tau$. The path-integral localizes on configurations of
$A,\phi,\sigma$ such that
\begin{align}
(F-\phi\wedge\phi)^+&=0, & (D\phi)^- &=0,\\
D\sigma &=0, & [\sigma,\bsigma]&=0,\\
[\phi,\sigma]&=0, & &
\end{align}
One obvious class of solutions of these equations is given by
anti-self-dual connections (instantons) with $\phi=\sigma=0$.
However, the instanton contribution to the partition function
vanishes because of fermionic zero modes, and in fact the partition
function is independent of $\tau$. Indeed, the partition function is
independent of $t$, we can use any of the BRST-operators $Q_t$ to
compute it, and the partition function for $t=i$ is
$\tau$-independent, as explained in \cite{KW}.

\section{Reduction to two dimensions}

Consider now the GL-twisted theory on a manifold of the form
$C\times\Sigma$, where $C$ and $\Sigma$ are Riemann surfaces. As
explained in \cite{KW}, for $t=0$ the effective field theory on
$\Sigma$ is the A-model whose target is the moduli space of Higgs
bundles on $C$, which we denote $\MH(G,C)$. The A-model depends only
on the K\"ahler form $\omega$ and B-field $B$ on $\MH(G,C)$ which
are given by
$$
B+i\omega=-\btau \omega_I,
$$
where
$$
\omega_I=-\frac{1}{4\pi}\int_C \Tr\, \left(\delta A\wedge\delta
A-\delta\phi\wedge\delta\phi\right).
$$
Note that $\omega_I$ is independent of the complex structure on $C$;
this is a reflection of the fact that the GL-twisted theory is a
topological field theory on $M=C\times\Sigma$.

As discussed in \cite{KW}, the moduli space $\MH(G,C)$ is
hyperk\"ahler, so it has three complex structures $I,J,K$ satisfying
$IJ=K$. The form $\omega_I$ is K\"ahler with respect to $I$. The
other two K\"ahler forms are
\begin{align*}
\omega_J & =\frac{i}{4\pi}\int_C dz\,d\bz\,
\Tr\,\left(\delta\phi_\bz\wedge
\delta A_z+\delta\phi_z\wedge \delta A_\bz\right),\\
\omega_K & =\frac{1}{2\pi}\int_C \Tr\, \delta\phi\wedge\delta A
\end{align*}
The form $\omega_J$ depends on the choice of complex structure on
$C$, while $\omega_K$ obviously does not, just like $\omega_I$. On
the other hand, both $\omega_J$ and $\omega_K$ are exact, while
$\omega_I$ is not; we normalized it so that its periods are integer
multiples of $2\pi$. In fact, the second cohomology group of
$\MH(G,C)$ is $\ZZ$, and the cohomology class of $\omega_I/2\pi$
generates it.

We note for future use that the form $\omega_I$ is cohomologous to
\begin{equation}\label{l0}
-\frac{1}{4\pi}\int_C \Tr\, \delta A\wedge\delta
A=-\frac{1}{2\pi}\int_C dz\, d\bz\, \Tr\, \delta A_z\wedge\delta
A_\bz.
\end{equation}
This form is a pull-back of a $(1,1)$ form on $\Bun_G(C)$ whose
periods are integer multiples of $2\pi$ and therefore can be thought
of as the curvature of a certain holomorphic line bundle $\cL_0$
over $\Bun_G(C)$. For $G=SU(N)$ this is simply the determinant line
bundle.

\section{From A-branes to noncommutative B-branes}

The Montonen-Olive duality of $N=4$ gauge theory implies that the
A-models with targets $\MH(G,C)$ and $\MH(\LG,C)$ are equivalent
provided the complexified K\"ahler classes are related by
\begin{equation}\label{Sdual:tau}
\Ltau=-\frac{1}{n_\frg\tau}.
\end{equation}
In mathematical terms, this means that the categories of A-branes
attached to $\MH(G,C)$ and $\MH(\LG,C)$ are equivalent. As explained
in \cite{KW}, one can understand this equivalence as coming from
T-dualizing the Hitchin fibration of $\MH(G,C)$. The Hitchin
fibration is holomorphic in complex structure $I$, therefore this
T-duality relates A-models in complex structure $I$ for $G$ and
$\LG$. Note that the same T-duality induces an equivalence of
B-models in complex structure $I$; as explained in \cite{Kap:HT},
this leads to the ``baby'' version of Geometric Langlands duality.

To make the statement of Montonen-Olive duality more useful, we
would like to relate A-branes in complex structure $I$ to some more
familiar mathematical objects. As in \cite{KW}, we can try to find a
special coisotropic A-brane analogous to the canonical coisotropic
A-brane of \cite{KW}. For $\Re\,\tau=0$, this is the A-brane given
by a rank-one line bundle on $\MH(G,C)$ with a connection whose
curvature is
$$
F=\Im\,\tau\,\omega_K.
$$
This form is exact:
$$
F=\Im\,\tau\, \frac{1}{2\pi}\delta\int_C \Tr\, \phi\wedge\delta A
$$
We will call this brane the distinguished coisotropic A-brane, or
d.c. brane for short, to distinguish it from the c.c. brane of
\cite{KW}. The c.c. brane has $F=\Im\,\tau\,\omega_J$. Note that the
definition of the d.c. brane is independent of the complex structure
on $C$, i.e. it is a topological object. On the other hand, the
definition of the c.c. brane involves the complex structure on $C$
in an essential way.

The space of boundary observables for the d.c. brane is the space of
functions on $\MH(G,C)$ holomorphic in complex structure
$J=\omega_I^{-1}\omega_K$. In this complex structure the space
$\MH(G,C)$ can be identified with the moduli space of stable flat
$G_\CC$ connections on $C$, which we denote $\cM_{flat}(G_\CC,C)$.
Furthermore, the algebra structure on the space of observables is
noncommutative: it is the quantization of the algebra of holomorphic
functions corresponding to the Poisson bivector
$$
P=(\Im\,\tau)^{-1}\Omega_J^{-1}.
$$
where
$$
\Omega_J=\omega_K+i\omega_I=-\frac{i}{4\pi}\int_C\Tr\,
\delta\cA\wedge\delta\cA =\frac{-i}{2\pi}\int_C dz\, d\bz\, \Tr\,
\delta\cA_z\wedge\delta\cA_\bz .
$$
More precisely, the action of the A-model on a disc has the form
$$
i\,\Im\,\tau\, \int \phi^*\Omega_J
$$
plus BRST exact terms. Thus $\Im\tau$ plays the role of the inverse
of the Planck constant. The algebra of boundary observables becomes
noncommutative, and to first order in $(\Im\tau)^{-1}$ we have
$$
[f,g]=-i(\Im\tau)^{-1}\Omega_J^{-1}(df,dg)+O((\Im\tau)^{-2}),
$$
where $f$ and $g$ are holomorphic functions on the moduli space of
flat $G_\CC$-connections. The functions $f$ and $g$ may only be
locally-defined, so we are really deforming the sheaf of holomorphic
functions on $\cM_{flat}(G_\CC,C)$.

To any A-brane $\alpha$ in complex structure $I$ we can attach a
module over this noncommutative algebra. This module is the space of
the states of the A-model on an interval with boundary conditions
given by $\alpha$ and the d.c. brane. More precisely, we can
associate to an A-brane a sheaf of modules over the sheaf of
boundary observables for the d.c. brane.

It has been argued in \cite{Kap:NC} (see also
\cite{Pestun,KW,GW:quant}) that whenever a coisotropic A-brane of
maximal dimension exists, the category of A-branes is equivalent to
the category of B-branes on a noncommutative deformation of the
underlying complex manifold. In the present case, this complex
manifold is the moduli space of stable flat $G_\CC$ connections on
$C$. Thus the Montonen-Olive duality can be reformulated as the
statement that the derived categories of coherent sheaves on the
quantizations of $\cM_{flat}(G_\CC,C)$ and $\cM_{flat}(\LG_\CC,C)$
are equivalent. As usual, we expect that the equivalence is given by
a certain object on the product of the quantized spaces. Its space
of sections gives rise to a bi-module over the algebras of quantized
functions on $\cM_{flat}(G_\CC,C)$ and $\cM_{flat}(\LG_\CC,C)$, i.e.
these algebras are derived-Morita-equivalent.

If $\Re\,\tau \neq 0$, then there is a B-field on $\MH(G,C)$ given
by
$$
B=-\Re\,\tau\, \omega_I
$$
If $\Re\,\tau=n\in\ZZ$, then there exists an analogue of the d.c.
brane which has the curvature
$$
F=\Im\,\tau\, \omega_I +  \Re\,\tau\, \omega_I.
$$
This curvature 2-form is not exact: its cohomology class is $n$
times the first Chern class of the determinant line bundle over
$\MH(G,C)$. The algebra of boundary observables for this brane is
the same as for $\Re\,\tau=0$. Indeed, this is clear from the fact
that the gauge theory for $\Re\,\tau=n$ is isomorphic to the gauge
theory for $\Re\,\tau=0$.

For more general values of $\Re\,\tau$ there is no rank-one
coisotropic A-brane on $\MH(G,C)$. However, for rational values of
$\Re\,\tau$ there may exist distinguished coisotropic A-branes of
higher rank. Indeed, the equation for the curvature of a $U(r)$
vector bundle on a coisotropic brane,
$$
F=\Im\,\tau\,\omega_K \cdot {\bf 1} + \Re\,\tau\, \omega_I \cdot
{\bf 1},
$$
may admit solutions  if $r \cdot\Re\,\tau$ is integral. The space of
boundary observables for such a brane is the space of holomorphic
sections of a certain algebra bundle which is locally isomorphic to
$\End(\eps)$ for some rank-$r$ holomorphic vector bundle $\eps$ on
$\MH(G,C)$. Such an algebra bundle is called an Azumaya algebra over
$\MH(G,C)$. The algebra of boundary observables is the quantization
of the algebra of sections of this Azumaya algebra. While there is
no canonical choice of such an Azumaya algebra, its category of
modules depends only on $\Re\,\tau$ (which determines the
Morita-class of the Azumaya algebra). Presumably, this remains true
after quantization. Note however that rationality of $\Re\,\tau$ is
not preserved by the Montonen-Olive duality.

Let us discuss some examples of A-branes and their conjectural duals
at $t=0$. First of all, we have the d.c. brane itself. Since its
curvature is of type $(1,1)$ in complex structure $K$, it is a brane
of type $(A,A,B)$ in the notation of \cite{KW}. That is, it is an
A-brane in complex structures $I$ and $J$ and is a B-brane with
respect to complex structure $K$. A slightly more general $(A,A,B)$
brane is obtained by twisting with a flat line bundle on $\MH(G,C)$.

The other obvious example is the c.c. brane of \cite{KW}, which is
obtained by letting
$$
F=\Im\,\tau\,\omega_J.
$$
This brane is of type $(A,B,A)$ and depends on the complex structure
on $C$. In fact, since the complex structures $J$ and $K$ are
related by an isometry, it can be obtained from the d.c. brane by a
hyperk\"ahler rotation.

A large class of Lagrangian A-branes is obtained by considering all
flat $G_\CC$-connections on $C$ with a fixed $(0,1)$ part, i.e.
fixed holomorphic structure. This defines a topologically trivial
Lagrangian submanifold of $\MH(G,C)$ which is moreover a complex
submanifold with respect to complex structure $J$. In other words,
it is an A-brane of type $(A,B,A)$. For example, one can require the
holomorphic structure to be trivial, or to require the flat
connection to be an oper on $C$. Obviously, one may also consider
all flat connections with a fixed $(1,0)$ part. This condition also
defines a complex Lagrangian submanifold with respect to complex
structure $J$, i.e. it is a brane of type $(A,B,A)$.

Our final example of a Lagrangian submanifold in complex structure
$I$ is inspired by instanton Floer homology. Consider a 3-manifold
$N$ whose boundary is the Riemann surface $C$. We define
$\alpha_N\subset \cM_{flat}(G_\CC,C)$ by saying that a point
$p\in\cM_{flat}(G_\CC,C)$ belongs to $\alpha_N$ if and only if the
corresponding flat connection is a restriction of a flat
$G_\CC$-connection on $N$. It is well-known that $\alpha_N$ is a
complex Lagrangian submanifold in $\cM_{flat}(G_\CC,C)$, i.e. it is
a brane of type $(A,B,A)$. Note that the condition defining
$\alpha_N$ is nonlocal on $C$. Thus while it is a valid A-brane for
a 2d sigma-model with target $\MH(G,C)$, it does not arise from a
local boundary condition in 4d gauge theory.

Let us make some remarks on the action of Montonen-Olive duality on
these A-branes. This duality maps branes of type $(A,B,A)$ to branes
of the same type. In particular, since the c.c. brane is flat along
the fibers of the Hitchin fibration, its dual must be a Lagrangian
$(A,B,A)$ brane. Furthermore, it must intersect each nonsingular
fiber of the Hitchin fibration at a single point. It is plausible
that the c.c. brane on $\MH(G,C)$ is dual to the submanifold of
opers in $\MH(\LG,C)$ \cite{KW}. Then the mirror of the d.c. brane
is a Lagrangian $(A,A,B)$ brane which is obtained from the oper
brane by the hyperk\"ahler rotation which turns $J$ into $K$.

\section{From noncommutative B-branes to twisted D-modules}

We have argued above that for $\theta=0$ the category of A-branes on
the moduli space of Higgs bundles is equivalent to the category of
B-branes on a noncommutative deformation of $\cM_{flat}(G_\CC,C)$.
We now want to reinterpret this result in terms of twisted D-modules
on $\Bun_G(C)$.

This reinterpretation is based on the fact that the moduli stack of
flat connections can be regarded as a twisted cotangent bundle over
$\Bun_G(C)$. This fact goes back to
\cite{BS,Falt}.\footnote{Strictly speaking, the space relevant for
us is the moduli space of stable flat connections, which is not a
cotangent bundle. Rather, an open subset of $\cM_{flat}(G,C)$ is a
cotangent bundle to the moduli space of stable $G$-bundles on $C$.
As a consequence, we will be really dealing with D-modules not on
$\Bun_G(C)$, but on the moduli space of stable $G$-bundles. We will
gloss over this important subtlety.} Let us recall what this means.

An affine bundle over a complex manifold $M$ modeled on $T^*M$ can
be described as follows. Let $q^i$ be local coordinate functions on
a coordinate chart $U\subset M$, and $\tq^i$ be local coordinate
functions on a coordinate chart $\tU$. Let $p_i$ and $\tp_i$ be the
corresponding ``Darboux'' coordinates on the fibers of $T^*U$ and
$T^*\tU$, respectively. The information about the affine bundle
$\bbA$ is contained in how $p_i$ is related to $\tp_i$ on
$U\bigcap\tU$:
\begin{equation}\label{ptransf}
\tp_i d\tq^i= p_i dq^i+\alpha_{U,\tU},
\end{equation}
where $\alpha_{U,\tU}$ is a holomorphic 1-form on $U\bigcap\tU$. It
is clear that the totality of these 1-forms is a 1-cocycle with
values in the sheaf $\Omega^1_M$, and that cohomologous cocycles
define isomorphic affine bundles. We will denote by $\alpha(\bbA)$
the class in $H^1(\Omega^1)$ corresponding to the affine bundle
$\bbA$.

If $\alpha(\bbA)$ is $\partial$-closed, the 2-form $dp_i dq^i$ on
the total space of $\bbA$ is well-defined and is holomorphic
symplectic. In such a case one says that $\bbA$ is a twisted
cotangent bundle over $M$. For example, if $\cL$ is a holomorphic
line bundle on $M$ equipped with a Hermitian metric, its curvature
gives a $\partial$-closed element in $H^1(\Omega^1_M)$. This
cohomology class does not really depend on the choice of metric. We
will denote this twisted cotangent bundle
$\bbA_\cL(M)$.\footnote{Another way to explain what $\bbA_\cL(M)$ is
is to say that it is the space associated to the sheaf of
holomorphic $\partial$-connections on $\cL$.} More generally, we can
multiply the curvature of $\nabla$ by a complex number $\lambda$;
the corresponding twisted cotangent bundle will be denoted
$\bbA_{\cL^\lambda}(M)$.

The basic fact we need is that the moduli space of flat
$G_\CC$-connections on $C$ equipped with the symplectic form
$\Omega_J$ is isomorphic to the twisted cotangent bundle $\bbA$ over
$\Bun_G(C)$ such that
$$
\alpha(\bbA)=\frac{i}{2\pi}\int_C dz\, d\bz\, \Tr \,\delta A_z\wedge
\delta A_\bz .
$$
The map to $\Bun_G(C)$ is the forgetful map which keeps only the
$(0,1)$ part of the connection. The fiber of the map is the space of
$\partial$-connections on a fixed holomorphic $G$-bundle, which is
an affine space modeled on the cotangent space to $\Bun_G(C)$. Since
the second cohomology of $\Bun_G(C)$ is one-dimensional and spanned
by the class of the $(1,1)$ form (\ref{l0}), the class $\alpha$ of
the resulting twisted cotangent bundle $\bbA$ must be proportional
to $(\ref{l0})$. To fix the proportionality constant we note that
the locally-defined ``holomorphic symplectic potential'' for the
holomorphic symplectic form $\Omega_J$ is
$$
-\frac{i}{2\pi}\int_C dz\, d\bz\, \Tr\, \cA_z\delta\cA_\bz .
$$
We can make it globally well-defined by adding to it a
locally-defined $(1,0)$ form
$$
-\frac{i}{2\pi}\int_C dz\, d\bz\, \Tr\,(-A_z+i\phi_z)\delta\cA_\bz .
$$
Acting on the resulting globally-defined $(1,0)$ form with
$\bpartial$, we get a $(1,1)$ form on $\Bun_G(C)$:
$$
\frac{-i}{2\pi}\int_C dz\, d\bz\, \Tr\,
\delta\cA_\bz^\dag\wedge\delta\cA_\bz .
$$
This $(1,1)$ form is cohomologous to
$$
\frac{i}{2\pi}\int_C dz\, d\bz\, \Tr\, \delta A_z \wedge\delta A_\bz
.
$$

We can formulate this result in a slightly different way by noting
that $\alpha(\bbA)$ is $-2\pi i$ times the first Chern class of the
line bundle $\cL_0$ on $\Bun_G(C)$. Then we can identify the moduli
space of flat connections with the twisted cotangent bundle
$\bbA_{\cL_0^{-1}}(\Bun_G)$.

The identification of the moduli space of flat $G_\CC$ connections
with a twisted cotangent bundle over $\Bun(G)$ enables one to
quantize it in a straightforward way. Let $M$ be a complex manifold,
let $\cL$ be a holomorphic line bundle over it, and let $\lambda$ be
a nonzero complex number. To quantize the structure sheaf of the
complex symplectic manifold $\bbA_{\cL^\lambda}(M)$ with the
symplectic form
$$
\frac{1}{\hbar}dp_j dq^j
$$
we impose the commutation relations
$$
[p_k,q^j]=-i\hbar\delta_k^j ,\quad [p_k, p_j]=0.
$$
We can try to solve these commutation relations by letting
$$
p_j=-i\hbar\nabla_j,
$$
where $\nabla_j$ is a (locally-defined) holomorphic differential on
a holomorphic line bundle $\cL_\hbar$ (or some complex power of a
holomorphic line bundle). Then the quantization of sheaf of
holomorphic functions can be identified with the sheaf of
holomorphic differential operators on $\cL_{\hbar}$. The line bundle
$\cL_\hbar$ is fixed by imposing the transformation law
(\ref{ptransf}). This implies $\cL_\hbar=\cL^{i\lambda/\hbar}$. Note
that the exponent $i\lambda/\hbar$ may be complex.

At this stage a subtlety creeps in. The first Chern class of
$\cL_\hbar$ that we got is of order $1/\hbar$; this happened because
the transformation law for $p_j$ was nontrivial already at the
classical level. One may wonder if there are quantum corrections to
the transformation law and therefore to $\cL_\hbar$. In fact, as
explained in \cite{KW}, the most natural quantization recipe leads
to such a correction. In the case of trivial $\cL$ it is natural to
take $\cL_\hbar=K^{1/2}$, where $K$ is the canonical line bundle of
$M$. This choice ensures that the algebra of twisted differential
operators is isomorphic to its opposite, something which is required
by a discrete symmetry $p\ra -p$ present for trivial $\cL$. From our
point of view, $K^{1/2}$ represents a correction to the first Chern
class of $\cL_\hbar$ which is of order $\hbar^0$. Asssuming that
there are no higher-order corrections, we conclude that the
quantization of $\bbA_\cL(M)$ is the sheaf of holomorphic
differential operators on $K^{1/2}\otimes\cL^{i\lambda/\hbar}$.

In our case $M=\Bun_G$, $\cL=\cL_0$, $\lambda=-1$, and
$\hbar=(\Im\,\tau)^{-1}$. We conclude that the sheaf of observables
for the d.c. brane can be identified with the sheaf of holomorphic
differential operators on
$K^{1/2}\otimes\cL_0^{-i\Im\tau}=\cL_0^{-h^\vee-i\Im\tau}$.
Therefore the Montonen-Olive duality at $t=0$ implies the statement
of the quantum geometric Langlands duality for $\qq=-i\Im\tau$. The
value of $\qq$ that we get in this way is purely imaginary.

\section{Line and surface operators at $t=0$}

\subsection{Line operators}

There are several different kinds of line operators. One kind of a
line operator is built using the descent procedure
\cite{Witten:top}. Given a local BRST-invariant 0-form $\cO$, one
may consider its descendant 1-form defined by
$$
d\cO = \delta \cO^{(1)}
$$
The integral of $\cO^{(1)}$ along a cycle $\gamma$ is then
BRST-invariant. For example, at $t=0$ we may take $\cO=\Tr\,
\sigma^2$, then
$$
\cO^{(1)}=\Tr\, \psi\sigma .
$$

Another kind of line operator is a Wilson-type line operator, i.e.
the holonomy of some connection. This connection must be
BRST-invariant up to a gauge transformation. At $t=0$ there are no
observables of this kind.

Finally, one may try to construct a disorder line operator. In the
present case, one needs to look for singular gauge field
configurations which solve the BPS equations. If we assume that near
the singularity the solutions become abelian, spherically symmetric
and scale-invariant, we can easily see that no disorder operators
can be constructed. Indeed, if the disorder operator is inserted at
$x^i=0,i=1,2,3$, then the vanishing of the BRST variation of
$\chi^+$ requires
$$
F_{ij}=\eps_{ijk}F_{4k}.
$$
Spherical symmetry and scale-invariance require
$$
F_{ij}=\frac{\rho}{2}\eps_{ijk} \frac{x^k}{r^3},
$$
where $\rho\in \frt$ lies in the cocharacter lattice of $G$ (this is
the Goddard-Nuyts-Olive quantization condition \cite{GNO}). But then
the ``electric field'' $F_{4k}$ is given by
$$
F_{4k}=\frac{\rho}{2} \frac{x^k}{r^3}.
$$
This is a Coulomb field, but the electric charge corresponding to it
is not quantized properly and in fact is purely imaginary rather
than real. Indeed, if we pass to Minkowski signature by replacing
$\partial_4\ra -i\partial_0$ and $F_{4k}\ra -iF_{0k}$, we find
$$
F_{0k}=\frac{i\rho}{2}\frac{x^k}{r^3}.
$$
Thus such a singularity does not correspond to a gauge-invariant
line operator.

\subsection{Surface operators}

Surface operators defined in \cite{GW:ramified} work equally well
for all $t$, including $t=0$ and $t=\infty$. The most basic surface
operators are parameterized by a quadruple
$(\alpha,\beta,\gamma,\eta)$, where $\beta,\gamma$ take values in
the Cartan subalgebra $\frt$ of $G$, $\alpha$ takes values in the
maximal torus $\bbT$ of $G$, and $\eta$ takes values in the maximal
torus ${}^L\bbT$ of $\LG$. In addition, one identifies quadruples
which differ by an element of the Weyl group and requires the
quadruple $(\alpha,\beta,\gamma,\eta)$ to be regular, in the sense
that no nontrivial element of the Weyl group leaves it invariant.
More generally, one can consider surface operators associated with a
Levi subgroup of $G$ \cite{GW:ramified}; the basic ones correspond
to taking $\bbT$ as the Levi subgroup.

At $t=0$ the parameters $\beta$ and $\gamma$ affect only the complex
structure of $\MH(G,C)$ and therefore locally are irrelevant as far
as the A-model is concerned.\footnote{Varying $\beta+i\gamma$ along
a closed loop in the allowed parameter space induces an
autoequivalence of the category of A-branes; these autoequivalences
play an important role in \cite{GW:ramified}.} The important
parameters are $\alpha$ and $\eta$ since they control the K\"ahler
form and the B-field on $\MH(G,C)$. Montonen-Olive duality acts on
them by
$$
(\alpha,\eta)\mapsto (\Lalpha,\Leta)=(\eta,-\alpha).
$$

We note in passing that although there are no nontrivial ``bulk''
line operators at $t=0$, line operators living on surface operators
seem to exist in some cases. For example, as discussed in
\cite{GW:ramified}, a noncontractible loop in the space of
``irrelevant'' parameters $\beta$ and $\gamma$ should correspond to
such a line operator. Such noncontractible loops exist if the pair
$(\alpha,\eta)$ is not regular, i.e. preserved by some nontrivial
element of the Weyl group.

\subsection{Ramified Quantum Geometric Langlands}

Surface operators affect the QGL story in the following way. Suppose
there are surface operators inserted at points $p_1,\ldots,p_s\in
C$. Then the effective 2d theory at $t=0$ is an A-model whose target
is the moduli space of stable parabolic Higgs bundles on $C$, with
parabolic structure at $p_1,\ldots,p_s$. We denote this moduli space
$\MH(G,C;p_1,\ldots,p_s)$. The K\"ahler form $\omega_I$ of
$\MH(G,C;p_1,\ldots,p_s)$ has periods which depend linearly on
$\Im\,\tau$ and $\alpha_1,\ldots,\alpha_s$; the B-field has periods
which depend linearly on $\Re\,\tau$ and $\eta_1,\ldots,\eta_s$.
More precisely, as explained in \cite{GW:ramified}, the second
cohomology of $\MH(G,C)$ can be naturally identified with
$$
\ZZ\oplus \oplus_{k=1}^s\Lambda_{char}(G),
$$
where $\Lambda_{char}(G)=\Hom(\bbT,U(1))\subset {}^L\frt$ is the
character lattice of $G$. In terms of this identification the
cohomology class of $\omega_I/2\pi$ is given by
$$
\left[\frac{\omega_I}{2\pi}\right]=e\oplus
\left(-\oplus_{k=1}^s\alpha^*_k\right),
$$
where the star denotes an isomorphism of $\frt$ and ${}^L\frt$
coming from the Killing form $-\Tr$. The K\"ahler form is
$\Im\,\tau\,\omega_I$. The cohomology class of the B-field is given
by \cite{GW:ramified}
$$
\left[\frac{B}{2\pi}\right]=\left(-\Re\,\tau\right)\oplus
\oplus_{k=1}^s\eta_k.
$$

Montonen-Olive duality now says that the categories of A-branes on
$\MH(G,C;p_1,\ldots,p_s)$ and $\MH(\LG,C;p_1,\ldots,p_s)$ are
equivalent if $\tau$ and $\Ltau$ are related as in (\ref{Sdual:tau})
and the parameters $(\alpha,\eta)$ and $(\Lalpha,\Leta)$ are related
by
\begin{equation}\label{Sdual:alphaeta}
\Lalpha_k=\eta_k,\quad \Leta_k=-\alpha_k,\quad k=1,\ldots,s.
\end{equation}

To connect this statement to a ramified version of QGL, we consider
the special case $\Re\,\tau=0$. We can try to define a d.c. brane by
requiring
$$
F+B=\Im\,\tau\, \omega_K
$$
for some cohomologically trivial 2-form $F$. This ensures that we
can interpret the 2-form $F$ as the curvature of a trivial line
bundle on $\MH(G,C)$. As explained in \cite{GW:ramified}, the
cohomology class of $\omega_K$ is given by
$$
\left[\frac{\omega_K}{2\pi}\right]=0\oplus \left(-\oplus_{k=1}^s
\gamma_k^*\right).
$$
Therefore the ``irrelevant'' parameters $\gamma_k$ are determined by
$\eta_k$:
$$
\gamma_k=-(\Im\,\tau)^{-1}\eta_k^*.
$$
This restriction is immaterial since the category of A-branes does
not depend on $\gamma$ anyway; we just need it to be able to
construct a d.c. brane.

Classically, the algebra of boundary observables for the d.c. brane
is the algebra of holomorphic functions on $\MH(G,C;p_1,\ldots,p_s)$
in complex structure $J$. The latter space can be identified with
the moduli space of stable parabolic $G_\CC$ local systems on $C$.

On the quantum level we need to quantize this algebra, or rather the
corresponding sheaf of algebras. As before, this is facilitated by
thinking about the moduli space of parabolic local systems as a
twisted cotangent bundle to the moduli space of parabolic
$G$-bundles on $C$. The natural holomorphic symplectic form
$\Omega_J=\omega_K+i\omega_I$ becomes the canonical symplectic form
on the twisted cotangent bundle. Therefore the quantized sheaf of
boundary observables is the sheaf of holomorphic differential
operators on $K^{1/2}\otimes \cL_\hbar$, where a priori $\cL_\hbar$
is a product of complex powers of line bundles over
$\Bun_G(C;p_1,\ldots,p_s)$. The first Chern class of $\cL_\hbar$ is
the cohomology class of $\Im\,\tau\,\Omega_J$ divided by $-2\pi $:
$$
c_1(\cL_\hbar)=i\,\Im\,\tau\, \left((-e)\oplus
\oplus_{k=1}^s\left(\alpha^*_k+i(\Im\,\tau)^{-1}\eta_k\right)\right).
$$
This class is no longer purely imaginary, but in the classical
regime (large $\Im\,\tau$) its imaginary part is much larger than
the real one. Note also that this class is a holomorphic function of
$\alpha^*_k+i(\Im\tau)^{-1}\eta_k$, as expected for the A-model in
complex structure $I$. The Montonen-Olive duality suggests that the
derived categories of twisted D-modules on
$\Bun_G(C;p_1,\ldots,p_s)$ and $\Bun_\LG(C;p_1,\ldots,p_s)$ are
equivalent if $\alpha_k,\eta_k$ and $\Lalpha_k$ and $\Leta_k$ are
related by (\ref{Sdual:alphaeta}) and $\Im\,\tau$ and $\Im\,\Ltau$
are related by
$$
\Im\,\Ltau=\frac{1}{n_\frg \Im\,\tau}
$$
This is a ramified version of the Quantum Geometric Langlands
duality.

\end{document}